\UseRawInputEncoding
\documentclass[%
reprint,
superscriptaddress,
nofootinbib,
 amsmath,amssymb,
 aps,
 prl,
floatfix,
]{revtex4-1}
\usepackage{graphicx}
\usepackage{subfigure}
\usepackage{dcolumn}
\usepackage{bm}
\usepackage{soul}
\usepackage{ulem}
\usepackage[%
bookmarksnumbered,
bookmarksopen,
colorlinks,
citecolor=blue,
linkcolor=blue,
]{hyperref} 

\def\es0{$E_{sym}(\rho_0)$~}
\def\us0{$U_{sym}^{\infty}(\rho_{0})$}
\begin{document}

\title {Proton Collectivity in Au+Au Collisions at $\sqrt{s_{\rm NN}}=2.4-4.5$~GeV from a Unified Purely Hadronic EOS without QCD Phase Transition}
 \author{Gao-Feng Wei}
 \email{Corresponding author: wei.gaofeng@gznu.edu.cn}
 \affiliation{School of Physics and Electronic Science, Guizhou Normal University, Guiyang 550025, China}
\author{Shuang-Jie Liu}
 \affiliation{School of Physics and Electronic Science, Guizhou Normal University, Guiyang 550025, China}
 \author{Yu-Liang Zhao}
 \affiliation{School of Physics and Electronic Science, Guizhou Normal University, Guiyang 550025, China}
 \author{Qi-Jun Zhi}
 \affiliation{Guizhou Provincial Key Laboratory of Radio Astronomy and Data Processing, Guizhou Normal University, Guiyang 550025, China}
 \author{Zhigang Xiao}
\email{Corresponding author: xiaozg@tsinghua.edu.cn}
\affiliation{Department of Physics, Tsinghua University, Beijing 100084, China}
\affiliation{Center for High Energy Physics, Tsinghua University, Beijing 100084, China}
\date{\today}
\begin{abstract}

The nuclear equation of state (EOS) is generally considered to soften in the density range of $2-5$ times the saturation density $\rho_0$.  Using a purely hadronic transport model, we calculate the  proton directed, sideward, and elliptic flows and their excitation functions in heavy-ion collisions (HICs) at $\sqrt{s_{\rm NN}}=2.4-4.5$~GeV  and compare with the HADES, E895, and STAR data. We find that a momentum-dependent mean field with a unified incompressibility $K_0=230$~MeV quantitatively  reproduces  the experimental proton flows up to 4.3 GeV,  at which the maximum density reaches approximately $5\rho_0$. At 4.5 GeV, however, the pure hadronic model fails to reproduce the proton directed and elliptic flow data,  providing circumstantial evidence for the onset of partonic degrees of freedom in HICs.  Our results provide a hadronic baseline to  characterize the high-density nuclear matter and to map the region of hadron-quark phase transition.  
\end{abstract}

\maketitle

\textit{Introduction~}{\bf --}
Collective flow in HICs provides a sensitive probe of the early-stage interaction dynamics and the degrees of freedom of the created matter. The number-of-constituent-quark (NCQ) scaling of elliptic flow has long been established as a signature of partonic collectivity in the quark-gluon plasma (QGP) created at top RHIC and LHC energies~\cite{ADAMS2005102,PhysRevLett.92.052302,Krzewicki_2011}. At lower beam energies, corresponding to higher baryon chemical potential, NCQ scaling is markedly violated at $\sqrt{s_{\mathrm{NN}}}=3.2$~GeV~\cite{STARPLB2022,2qhx-cp79}, in accordance with the picture of a hadron-dominated phase, 
 but  reappears at $\sqrt{s_{\mathrm{NN}}}=4.5$~GeV, signaling the onset of dominant partonic interactions~\cite{2qhx-cp79,STAR21NCQ}.
 This beam-energy dependence offers an effective means to map the transition region between hadronic and quark matter on the QCD phase diagram. However, whether  the phase transition is sharp or extended, and whether its location depends on the observable,  remain open questions that  demand a rigorous and quantitative test with a  well-established hadronic baseline.

A decisive test relies on the reproduction of collective-flow data using a purely hadronic transport model with a global, beam-energy-independent EOS across the wide density range of $(2-5)\rho_0$. However, such a unified hadronic EOS has not yet been achieved, despite its critical importance for establishing the baseline against which genuine partonic signatures can be identified. Early attempts by the E895 Collaboration~\cite{Pinkenburg99,liu00}  have suggested that the nuclear EOS favors a hard form  at low densities and softens at high densities, with the transition occurring at about $4\rho_0$ achieved in Au+Au collisions at $\sqrt{s_{\mathrm{NN}}}=3.3$~GeV.  Moreover, combining directed and elliptic flow, a wide range of incompressibility  $K_0$ from $167$ to $380$~MeV was reported~\cite{Danielewicz02}.

Momentum-dependent interactions (MDI) are known to play a significant role in modeling HICs, regardless of the specific model applied~\cite{Dh23,Liu26,Zhou25,HADES20,STAR22}. Nevertheless, the theoretical predicament in reproducing flow data over a wide energy range persists even with the introduction of MDI. Model predictions exhibit strong energy-dependent inconsistencies when confronted with directed and elliptic flow data. For instance, it  has been reported  that a soft momentum-dependent EOS best describes low-energy collective flow data, while a hard  one is required at high energies~\cite{Tara2024}, clearly showing the demand for an energy-uniform description. Moreover, existing transport codes  often yield divergent predictions for the same flow observable even  with similar EOS inputs, as reported by recent systematic model comparisons~\cite{PhysRevC.93.044609,Kolo2005}. This persistent predicament highlights the theoretical difficulty of describing the pressure generated by compressing nuclear matter in the critical $(2-5)\rho_0$ region accessible in HICs, and underscores the need for further improvements in modeling the collision process. Among these,  careful treatments of the main inelastic channels are essential, since they affect nucleon-nucleon (NN) collisions and  consequently the flow behavior.

\begin{figure*}[tbp]
	\centering
	\includegraphics[width=1.5\columnwidth]{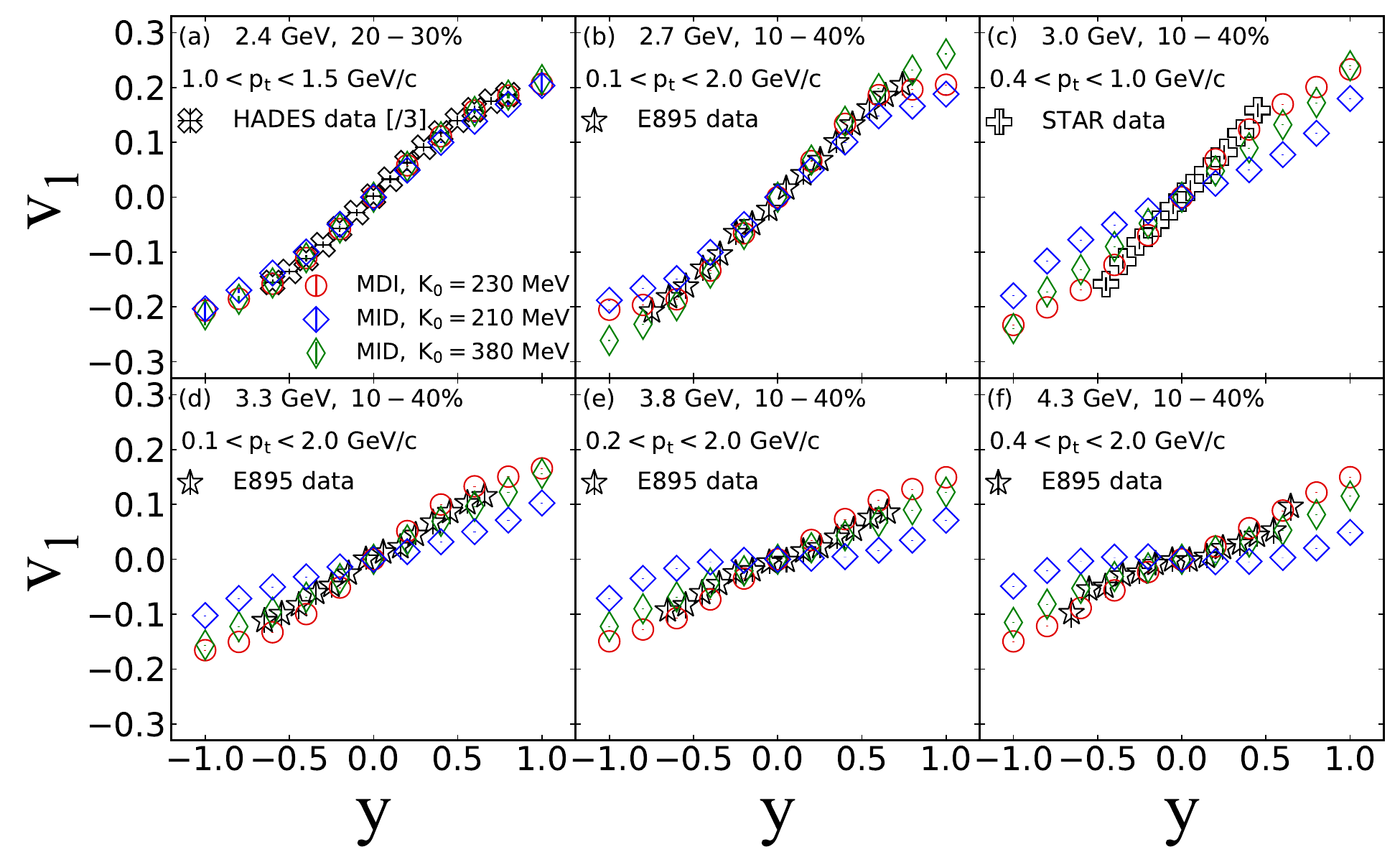}
	\caption{The directed flow of free protons as a function of rapidity in center of mass frame in Au+Au collisions at $\sqrt{s_{\rm NN}}=2.4-4.3$~GeV. The corresponding HADES, E895, and STAR data are taken from Refs.~\cite{HADES20,Pinkenburg99,liu00,STAR22}.  HADES data points (a), of both simulation and experiment, are divided by 3 for a convenient display in the same ordinate range as other panels. } 
	\label{v1y}
\end{figure*}

In this Letter, we address this long-standing problem by simulating the proton collective flow in Au+Au collisions from $\sqrt{s_{\mathrm{NN}}}=2.4$ to $4.5$~GeV using a purely hadronic transport model. We carefully treat the inelastic NN channels, including   $\Delta$ and $N^*$ resonances, strange baryons, and multi-pion production channels. Different scenarios of the EOS are compared.  A momentum-dependent mean field with $K_0=230$~MeV achieves global descriptions of the directed, sideward, and elliptic flows of protons up to $\sqrt{s_{\mathrm{NN}}}=4.3$~GeV,  corresponding to a maximum density of about $5\rho_0$, without   density-dependent softening. In  contrast, momentum-independent mean fields fail to reproduce the data. Remarkably, at $\sqrt{s_{\mathrm{NN}}}=4.5$~GeV, using the same momentum dependent EOS  fails to reproduce the proton flow data, despite its success below 4.3 GeV. This is the same energy at which the STAR Collaboration observed the restoration of NCQ scaling~\cite{2qhx-cp79,STAR21NCQ}.  Such a success-and-failure pattern of the purely hadronic simulations thus has important implications  for the hadron-quark phase transition.

\textit{The Model~}{\bf --}
The study is carried out within an isospin-dependent BUU (IBUU) transport model. To perform the comparative study, we incorporate two types of nuclear mean  field scenarios. 

The first is a default momentum-dependent nuclear mean field (labelled as MDI)~\cite{Das03,li04,Chen05} written as,
\begin{eqnarray}
	U(\rho,\delta ,\vec{p},\tau ) &=&A_{u}\frac{\rho _{-\tau }}{\rho _{0}}%
	+A_{l}\frac{\rho _{\tau }}{\rho _{0}}\notag \\
	&+&B{\Big(}\frac{\rho}{\rho _{0}}{\Big)}^{\sigma }(1-x\delta^2)-4\tau x\frac{B}{\sigma+1}\frac{\rho^{\sigma-1}}{\rho_{0}^\sigma}\delta\rho_{-\tau}
	\notag \\
	&+&\frac{2C_{l }}{\rho _{0}}\int d^{3}p^{\prime }\frac{f_{\tau }(%
		\vec{p}^{\prime })}{1+(\vec{p}-\vec{p}^{\prime })^{2}/\Lambda ^{2}}
	\notag \\
	&+&\frac{2C_{u }}{\rho _{0}}\int d^{3}p^{\prime }\frac{f_{-\tau }(%
		\vec{p}^{\prime })}{1+(\vec{p}-\vec{p}^{\prime })^{2}/\Lambda ^{2}},
	\label{IMDIU}
\end{eqnarray}%
where $\tau=1$ for neutrons and $-1$ for protons, and $A_{u}$, $A_{l}$, $C_{u}(\equiv C_{\tau,-\tau})$ and $C_{l}(\equiv C_{\tau,\tau})$ are connected to the properties of nuclear matter at $\rho_{0}=0.16$~fm$^{-3}$, 
and the parameter $x$ is used to mimic the slope value $L \equiv 3\rho_0 \, (dE_{\rm sym}/d\rho)|_{\rho_0}$ of symmetry energy at $\rho_{0}$. To examine the stiffness of nuclear  isoscalar EOS, we adopt $x=0$,  which corresponds to $L=62.5$~MeV. Other parameters are determined by fitting experimental/empirical constraints on properties of nuclear matter at $\rho_{0}$. Here we take the incompressibility  $K_{0}=230$~MeV (marked as MDI-230) for symmetric nuclear matter at $\rho_{0}$.  Further details can be found in Ref.~\cite{Liu26}.

The second scenario of the mean field is the momentum-independent interaction (labelled as MID)~\cite{Bertsch88,Baran05,Li08},  which is given by
\begin{eqnarray}	
U(\rho,\delta,\tau)=\alpha\left(\frac{\rho}{\rho_0}\right)+\beta\left(\frac{\rho}{\rho_0}\right)^\xi+\nu_{asy}(\rho,\delta,\tau),
\end{eqnarray} 
 where the third term,  $\nu_{asy}(\rho, \delta, \tau )$ \cite{Li02}, takes the form
\begin{equation}\begin{aligned}
		\nu_{asy}(\rho,\delta,\tau) & =\tau\left(E_{sym}(\rho_{0})u^{\gamma}-12.7u^{2/3}\right)\delta \\
		& +\left(E_{sym}(\rho_{0})(\gamma-1)u^{\gamma}+4.2u^{2/3}\right)\delta^{2},
\end{aligned}\end{equation}
describes  the isovector potential leading to the symmetry energy in the form $E_{sym}(\rho)=E_{sym}(\rho_0)u^{\gamma}$,
 with $u\equiv\rho/\rho_{0}$ being the reduced density. The parameter $\gamma$  controls the stiffness of $E_{sym}(\rho)$. For consistency,  we set $\gamma$=0.64, corresponding to  $L=62.5$ MeV in the MID scenario. The remaining parameters $\alpha$, $\beta$ and $\xi$ determine the incompressibility of the nuclear EOS. For  the comparative study, we adopt two parameter sets. A stiff EOS with $K_0=380$~MeV (marked as MID-stiff) corresponds to   $\alpha=-123.03$ MeV, $\beta=69.77$~MeV and $\xi=2.01$, and a soft EOS with $K_0=210$~MeV (marked as MID-soft), corresponds to  $\alpha=-303.32$~MeV, $\beta=250.06$~MeV and $\xi=1.21$.

\begin{figure}[tbp]
	\centering
	\includegraphics[width=0.75\columnwidth]{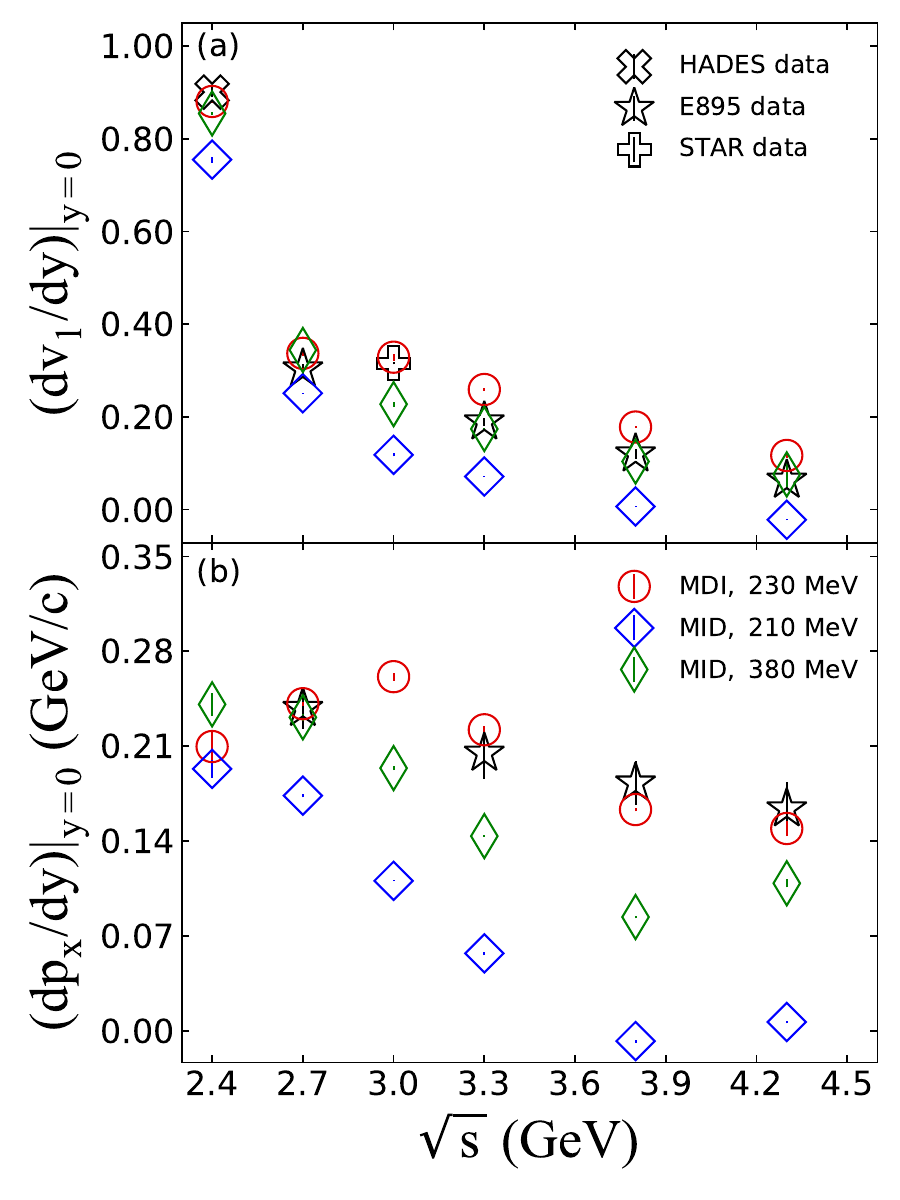}
	\caption{Excitation functions of proton directed flow $dv_1/dy|_{y=0}$ (a) and sideward flow $dp_x/dy|_{y=0}$ (b), in comparison to HADES~\cite{HADES20}, E895~\cite{Pinkenburg99,liu00}  and STAR ~\cite{STAR22} experiments.}
	\label{slope}	
\end{figure}

To simulate HICs at the investigated energies, we  extend the IBUU model with additional channels for baryon resonance and meson production, as in ART~\cite{Li95} and AMPT~\cite{Lin05}, among others. The propagating particles in the model include neutron ($n$), proton ($p$), $\Delta(1232)$, $N^*(1440)$, $N^*(1535)$, $\pi$, $K$, $\eta$, $\Lambda$ and $\Sigma$  etc. The corresponding cross sections  are taken from fits to the  experimental/empirical data~\cite{PDG88,Olive14}, as applied in our recent studies~\cite{Liu26,Wei24}.  We also include multi-pion production and the associated $\rho$ and $\omega$ channels, which contribute significantly, for instance, accounting for more than 30\% of the total $pp$ cross section at $p_{\rm lab}=6$~GeV/c. Technical details and additional discussions on the inclusion of these inelastic channels are provided in the Supplemental Material.

\begin{figure}[tbp]
	\centering
	\includegraphics[width=0.77\columnwidth]{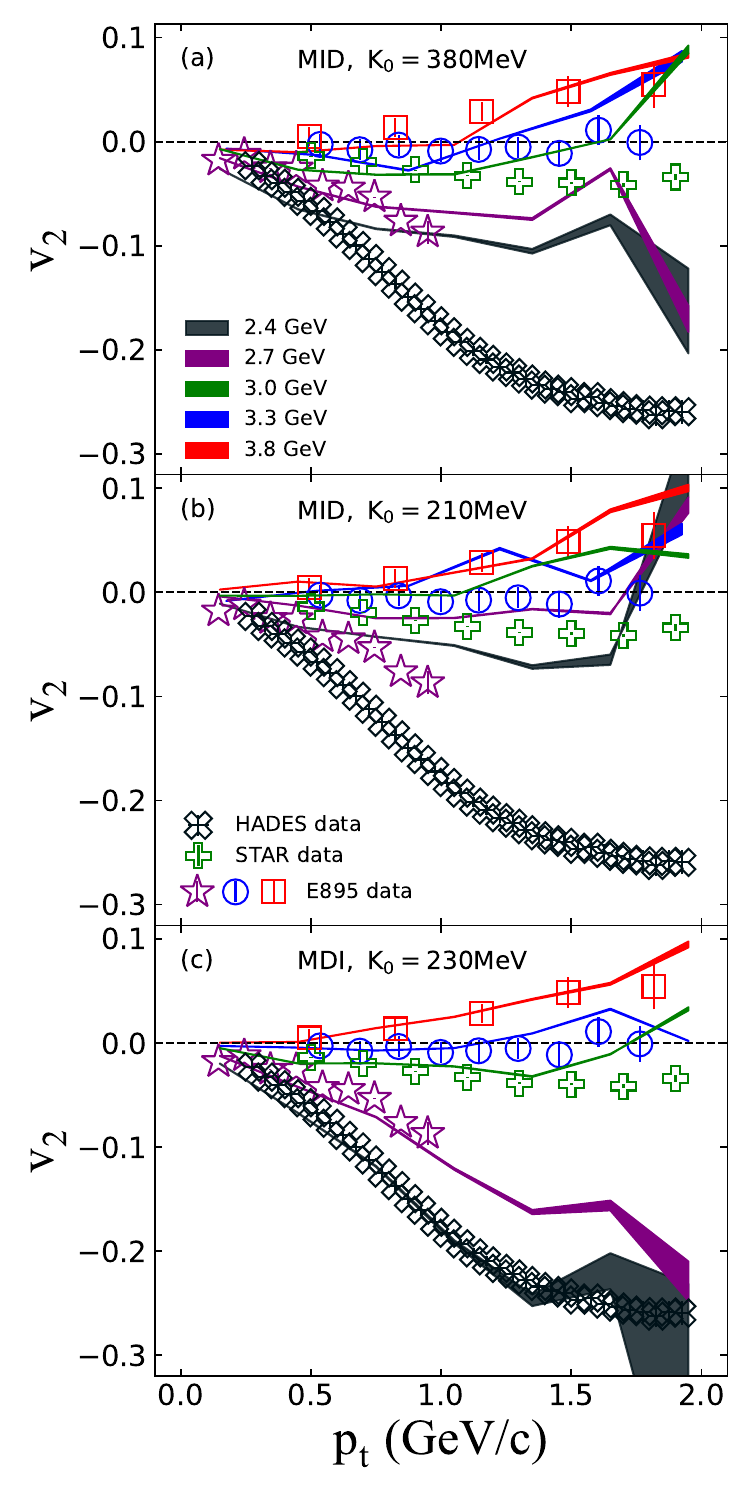}
	\caption{Proton elliptic flow as a function of transverse momentum in Au+Au collisions at $\sqrt{s_{\rm NN}}=2.4-3.8$~GeV with MID-stiff (a), MID-soft (b) and MDI-230 (c) potentials, respectively, in comparison to HADES~\cite{HADES20}, E895~\cite{Pinkenburg99,liu00}  and STAR ~\cite{STAR22} experiments.}
	\label{v2pt}
\end{figure}

\textit{Results and Discussions~}{\bf --}
We first present the proton flow observable in Au+Au collisions at $\sqrt{s_{\rm NN}}=2.4-4.3$~GeV. The corresponding results at $\sqrt{s_{\mathrm{NN}}}=4.5$~GeV will be shown later. Figure~\ref{v1y}  shows the rapidity dependence of the directed flow of free protons 
for two mean-field scenarios,  together with the corresponding data from HADES~\cite{HADES20}, E895~\cite{Pinkenburg99,liu00}, and STAR~\cite{STAR22}. All the calculations are filtered  with the same acceptance cuts and centrality conditions as used in the corresponding experiments. Free protons are identified  using the density criterion ($\rho<\rho_0/8$) and/or  the relative distance condition  in phase space ($\Delta_{R}>3.575$~fm or $\Delta_{p}>0.3$~GeV/$c$), as  in our recent study~\cite{Long24} in simulation of FOPI flow data~\cite{FOPI12}, and also as in coalescence models~\cite{Hillm20,Nagle96}. Clearly, both the MDI-230 and MID-stiff potentials  reproduce the proton directed flow data,  consistent with previous observations at low energies~\cite{Pan93,Zhang94}. Moreover, due to the stronger pressure and/or pressure gradient, the MID-stiff case yields a larger directed-flow amplitude than the MID-soft case. As sideward flow shares similar characteristics with directed flow,  the MDI-230 and MID-stiff mean-field potentials also describe the proton sideward flow quite well, as illustrated in the Supplemental Material.

\begin{figure}[tbp]
	\centering
	\includegraphics[width=0.77\columnwidth]{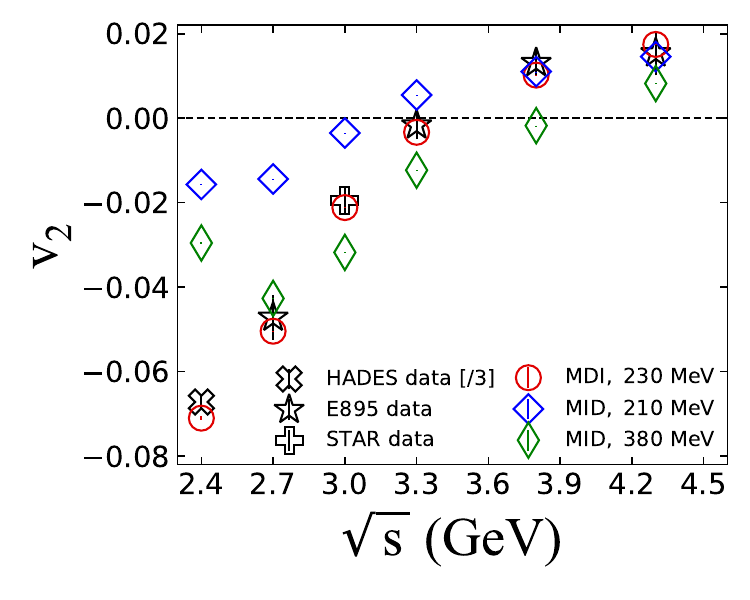}
	\caption{Excitation function of proton elliptic flow $v_2$ at the midrapidity, compared to  HADES~\cite{HADES20}, E895~\cite{Pinkenburg99,liu00}  and STAR ~\cite{STAR22} experiments. HADES data points, of both simulation and experiment, are divided by 3 for a convenient display in the same ordinate range.}
	\label{v2}
\end{figure}

 Figure~\ref{slope} presents the excitation functions of proton directed (a) and sideward (b) flows up to 4.3 GeV, compared with experimental data, to illustrate the effects of the pressure and/or pressure gradient from different nuclear EOS and momentum dependence of the mean field. The MDI-230 calculations agree reasonably with both excitation functions. The MID-stiff EOS reproduces the excitation functions for the directed flow but fails for the sideward flow. By contrast, the MID-soft EOS underestimates both excitation functions.

 Figure~\ref{v2pt} compares the proton elliptic flow as a function of transverse momentum ($p_t$) at  $\sqrt{s_{\rm NN}}=2.4-4.3$~GeV from the MID-stiff (a), MID-soft (b), and MDI-230 (c) potentials with experimental data.   None of the MID calculations reproduce the $p_t$ dependence of $v_2$, though MID-stiff case performs slightly better at low $p_t$ than MID-soft case.  In contrast, the MDI-230 calculations describe the $p_t$ dependence of $v_2$ for all available data. Notably, the nearly constant $v_2 \approx 0$ at $\sqrt{s_{\rm NN}}=3.3$~GeV indicates that $v_{2}$ changes sign around this energy.

Figure~\ref{v2} further compares the excitation function of the elliptic flow at midrapidity for Au+Au collisions at $\sqrt{s_{\rm NN}}=2.4-4.3$~GeV  from the MDI-230, MID-stiff, and MID-soft potentials  with experimental data. The simulations capture the observed sign change of $v_{2}$ from negative to positive with the increase of collision energy.  Quantitatively, however, the calculations with MID-stiff and MID-soft mean fields both deviate from the experimental data and  fail to reproduce the correct transition energy.  A comparison with the MID calculations suggests that the data favor a softening of the nuclear EOS, consistent with  earlier  conclusions~\cite{Pinkenburg99,liu00}. By contrast, the MDI-230 calculations  reproduce both the experimental excitation function of proton elliptic flow and the transition occurrence  around $\sqrt{s_{\rm NN}}=3.3$~GeV, confirming an energy-uniform description of the stiffness of nuclear matter without softening and highlighting the importance of the momentum dependent potential.  

\begin{figure}[tbp]
	\centering
	\includegraphics[width=0.75\columnwidth]{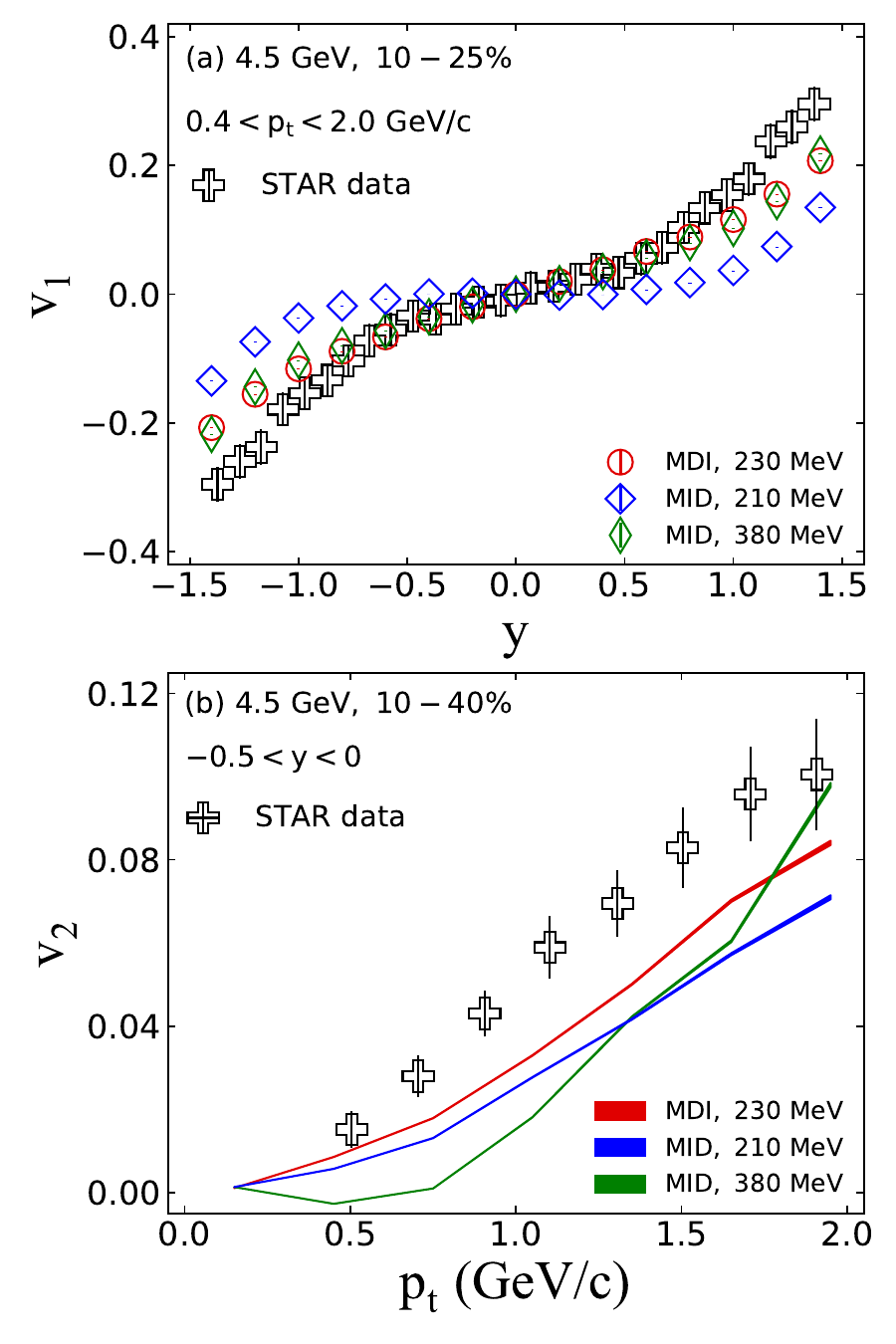}
	\caption{Proton directed flow as a function of rapidity in center of mass frame in Au+Au collisions at $\sqrt{s_{\rm NN}}=4.5$~GeV (a). Proton elliptic flow as a function of transverse momentum in the same  collisions (b). The corresponding STAR data are taken from Refs.~\cite{STAR21NCQ,2qhx-cp79}.}
	\label{v2-4.5}
\end{figure}

  We now examine the model predictions at $\sqrt{s_{\mathrm{NN}}}=4.5$~GeV. Figure~\ref{v2-4.5} compares the model predictions with different EOS to the experimental data for the proton  $v_{1}(y)$ (a) and $v_{2}(p_{t})$ (b). The model predictions  systematically deviate from the data for both $v_1$ and $v_2$, in strong contrast to the success achieved at  $\sqrt{s_{\mathrm{NN}}}=2.4-4.3$~GeV. Quantitatively, the $\chi^2/{\rm DoF}$  for the MDI-230 EOS  increases from 0.98 (Fig. \ref{v2}) to 5.2 (Fig. \ref{v2-4.5} (b)). Thus, the purely hadronic description,  successful up to $4.3$~GeV, breaks down at $4.5$~GeV.

The success of our purely hadronic MDI-230 EOS in globally reproducing the directed, sideward, and elliptic flow data across $\sqrt{s_{\mathrm{NN}}}=2.4-4.3$~GeV, together with its failure at 4.5 GeV, carries twofold implications for the QCD phase diagram at high baryon chemical potential.

First, our purely hadronic EOS requires no density-dependent softening in the  domain of $(2-5)\rho_0$.  The pressure of compressed nuclear matter in HICs up to  4.3 GeV is accurately captured by a hadronic description with a momentum-dependent interaction. This finding is at variance with earlier  inferences that the EOS softens at $4\rho_0$. 

Second,  the same model fails to reproduce the proton flow data   at $\sqrt{s_{\mathrm{NN}}}=4.5$~GeV, indicating  a breakdown of the purely hadronic description.  This sudden failure across a 0.2~GeV energy gap places a stringent constraint for the onset of a possible first-order transition, narrowing it to the window $4.3 < \sqrt{s_{\mathrm{NN}}} \le 4.5$~GeV, where the maximum density reaches slightly above $5\rho_0$. This result coincides with the  observed onset of NCQ scaling restoration at 4.5 GeV, a nearly model-independent signature of partonic collectivity reported by STAR Collaboration.

\textit{Conclusion~}{\bf --} 
We confront the IBUU transport model with the full set of proton directed, sideward, and elliptic flow data from the HADES, E895, and STAR collaborations across $\sqrt{s_{\rm NN}}=2.4-4.5$~GeV.   The momentum-dependent mean-field potential with  $K_0=230$~MeV quantitatively reproduces the experimental  proton flows up to $4.3$~GeV. 
This implies that the nuclear EOS undergoes no softening over the broad density range $(2-5)\rho_0$, with the incompressibility remaining uniform  at the characteristic value. Such a uniform hadronic EOS  supports the existence of massive ($\gtrsim 2M_\odot$) neutron stars \cite{LiAng_2021,Li_2026}. However, the same purely hadronic model fails to describe the proton flow data at $4.5$~GeV, signaling the breakdown of the hadronic picture. Our results thus establish a robust, data-driven baseline for characterizing the high-density nuclear-matter EOS while pointing to the possible onset of a hadron-to-quark phase transition around $\sqrt{s_{\rm NN}}=4.5$~GeV. 

\textit{Acknowledgments~}{\bf --}
We thank Professors Bao-An Li, Nu Xu, Peng-Fei Zhuang, Shu-Su Shi, Ang Li, and Gao-Chan Yong for carefully reading the manuscript and providing valuable suggestions. 
This work is supported by the National Natural Science Foundation of China under grant Nos. 12475131, 12335008 and 11927901,  and the project of Guizhou Provincial Department of Science and Technology under Grant No. ZD[2026]108. ZGX acknowledges the support from the Initiative Scientific Research Program of Tsinghua University. 

\bibliography{reference.bib}

@article{ADAMS2005102,
title = {Experimental and theoretical challenges in the search for the quark–gluon plasma: The STAR Collaboration's critical assessment of the evidence from RHIC collisions},
author = {J. Adams and others},
collaboration = {STAR Collaboration},
journal = {Nucl. Phys. A},
volume = {757},
pages = {102-183},
year = {2005},
url = {https://doi.org/10.1016/j.nuclphysa.2005.03.085}
}

@article{PhysRevLett.92.052302,
title = {Particle-Type Dependence of Azimuthal Anisotropy and Nuclear Modification of Particle Production in $\mathrm{A}\mathrm{u}+\mathrm{A}\mathrm{u}$ Collisions at $\sqrt{{s}_{NN}}=200\text{ }\text{ }\mathrm{G}\mathrm{e}\mathrm{V}$},
author = {J. Adams and others},
collaboration = {STAR Collaboration},
journal = {Phys. Rev. Lett.},
volume = {92},
pages = {052302},
year = {2004},
url = {https://doi.org/10.1103/PhysRevLett.92.052302}
}

@article{Krzewicki_2011,
title = {Elliptic and triangular flow of identified particles measured with the ALICE detector at the LHC},
author = {Krzewicki, Mikołaj and (for the ALICE Collaboration)},
journal = {J. Phys. G: Nucl. Part. Phys.},
volume = {38},
pages = {124047},
year = {2011},
url = {https://doi.org/10.1088/0954-3899/38/12/124047}
}

@article{STARPLB2022,
title = {Disappearance of partonic collectivity in sNN=3GeV Au+Au collisions at RHIC},
journal = {Phys. Lett. B},
author = {M.S. Abdallah and others},
collaboration = {STAR Collaboration},
volume = {827},
pages = {137003},
year = {2022},
url = {https://doi.org/10.1016/j.physletb.2022.137003}
}

@article{2qhx-cp79,
title = {Onset of Constituent Quark Number Scaling in Heavy-Ion Collisions at RHIC},
author = {Aboona, B. E. and  others},
collaboration = {STAR Collaboration},
journal = {Phys. Rev. Lett.},
volume = {135},
pages = {072301},
year = {2025},
url = {https://doi.org/10.1103/2qhx-cp79}
}

@article{STAR21NCQ,
title = {Flow and interferometry results from Au + Au collisions at sNN = 4.5 GeV},
author = {M. S. Abdallah and others},
collaboration = {STAR Collaboration},
journal = {Phys. Rev. C},
volume = {103},
pages = {034908},
year = {2021},
url = {https://doi.org/10.1103/PhysRevC.103.034908}
}

@article{Pinkenburg99,
title = {Elliptic flow: transition from out-of-plane to in-plane emission in Au + Au collisions},
author = {C. Pinkenburg and N. N. Ajitanand and J. M. Alexander and others},
collaboration = {E895 Collaboration},
journal = {Phys. Rev. Lett.},
volume = {83},
pages = {1295},
year = {1999},
url = {https://doi.org/10.1103/PhysRevLett.83.1295}
}

@article{liu00,
title = {Sideward flow in Au+Au collisions between 2A and 8A GeV},
author = {H. Liu and N. N. Ajitanand and J. Alexander and others},
collaboration = {E895 Collaboration},
journal = {Phys. Rev. Lett.},
volume = {84},
pages = {5488},
year = {2000},
url = {https://doi.org/10.1103/PhysRevLett.84.5488}
}

@article{Danielewicz02,
title = {Determination of the equation of state of dense matter},
author = {P. Danielewicz and R. Lacey and W.G. Lynch},
journal = {Science},
volume = {298},
pages = {1592},
year = {2002},
url = {https://doi.org/10.1126/science.1078070}
}

@article{Dh23,
title = {Directed and elliptic flows of protons and deuterons in HADES Au+Au collisions at sNN=2.4 GeV},
author = {Huan Du and G. F. Wei and G. C. Yong},
journal = {Phys. Lett. B},
volume = {839},
pages = {137823},
year = {2023},
url = {https://doi.org/10.1016/j.physletb.2023.137823},
}

@article{Liu26,
title = {Proton and kaon production in Au+Au collisions at sNN = 3 GeV},
author = {S. J. Liu and G. F. Wei and Y. L. Zhao and F. C. Zhou and Z. Wang},
journal = {Nucl. Sci. Tech.},
note = {in press},
year = {2026},
eprint = {2604.27678},
archivePrefix = {arXiv}
}

@article{Zhou25,
title = {Probing the nuclear equation of state with cluster and hypernuclei},
author = {Y. J. Zhou and S. Gl\"{a}ssel and Y. H. Leung and others},
journal = {Phys. Rev. C},
volume = {113},
pages = {014909},
year = {2026},
url = {https://doi.org/10.1103/3msm-wrxd}
}

@article{HADES20,
title = {Directed, elliptic, and higher order flow harmonics of protons, deuterons, amd tritons in Au+Au collisions at $\sqrt{s_{\mathrm{NN}}}=2.4$\,GeV},
author = {J. Adamczewski-Musch and others},
collaboration = {HADES Collaboration},
journal = {Phys. Rev. Lett.},
volume = {125},
pages = {262301},
year = {2020},
url = {https://doi.org/10.1103/PhysRevLett.125.262301}
}

@article{STAR22,
title = {Light nuclei collectivity from sNN = 3 GeV Au+Au collisions at RHIC},
author = {M. S. Abdallah and others},
collaboration = {STAR Collaboration},
journal = {Phys. Lett. B},
volume = {827},
pages = {136941},
year = {2022},
url = {https://doi.org/10.1016/j.physletb.2022.136941}
}

@article{Tara2024,
title = {Flow and equation of state of nuclear matter at $E_{\mathrm{kin}/A}=0.25$–1.5 GeV with the SMASH transport approach},
author = {L. A. Tarasovi\v{c}ov\'{a} and J. Mohs and A. Andronic and H. Elfner and K. H. Kampert},
journal = {Eur. Phys. J. A},
volume = {60},
pages = {232},
year = {2024},
url = {https://doi.org/10.1140/epja/s10050-024-01445-w}
}

@article{PhysRevC.93.044609,
title = {Understanding transport simulations of heavy-ion collisions at $100A$ and $400A$ MeV: Comparison of heavy-ion transport codes under controlled conditions},
author = {Xu, Jun and Chen, Lie Wen and Tsang, ManYee Betty and Wolter, Hermann and Zhang, Ying Xun and Aichelin, Joerg and Colonna, Maria and Cozma, Dan and Danielewicz, Pawel and Feng, Zhao Qing and Le F\`evre, Arnaud and Gaitanos, Theodoros and Hartnack, Christoph and Kim, Kyungil and Kim, Youngman and Ko, Che Ming and Li, Bao An and Li, Qing Feng and Li, Zhu Xia and Napolitani, Paolo and Ono, Akira and Papa, Massimo and Song, Taesoo and Su, Jun and Tian, Jun Long and Wang, Ning and Wang, Yong Jia and Weil, Janus and Xie, Wen Jie and Zhang, Feng Shou and Zhang, Guo Qiang},
journal = {Phys. Rev. C},
volume = {93},
pages = {044609},
year = {2016},
url = {https://doi.org/10.1103/PhysRevC.93.044609}
}

@article{Kolo2005,
title = {Transport theories for heavy-ion collisions in the 1 A GeV regime},
author = {Kolomeitsev, E E and Hartnack, C and Barz, H W and Bleicher, M and Bratkovskaya, E and Cassing, W and Chen, L W and Danielewicz, P and Fuchs, C and Gaitanos, T and Ko, C M and Larionov, A and Reiter, M and Wolf, Gy and Aichelin, J},
journal = {J. Phys. G: Nucl. Part. Phys.},
volume = {31},
pages = {S741},
year = {2005},
url = {https://doi.org/10.1088/0954-3899/31/6/015}
}

@article{Das03,
title = {Momentum dependence of symmetry potential in asymmetric nuclear matter for transport model calculations},
author = {C. B. Das and S. Das Gupta and C. Gale and B. A. Li},
journal = {Phys. Rev. C},
volume = {67},
pages = {034611},
year = {2003},
url = {https://doi.org/10.1103/PhysRevC.67.034611}
}

@article{li04,
title = {Momentum dependence of the symmetry potential and nuclear reactions induced by neutron-rich nuclei at RIA},
author = {B. A. Li and C. B. Das and S. Das Gupta and C. Gale},
journal = {Phys. Rev. C},
volume = {69},
pages = {011603(R)},
year = {2004},
url = {https://doi.org/10.1103/PhysRevC.69.011603}
}

@article{Chen05,
title = {Determination of the stiffness of the nuclear symmetry energy from isospin diffusion},
author = {L. W. Chen and C. M. Ko and B. A. Li},
journal = {Phys. Rev. Lett.},
volume = {94},
pages = {032701},
year = {2005},
url = {https://doi.org/10.1103/PhysRevLett.94.032701}
}

@article{Bertsch88,
title = {A guide to microscopic models for intermediate energy heavy ion collisions},
author = {G.F. Bertsch and S. Das Gupta},
journal = {Phys. Rep.},
volume = {160},
pages = {189},
year = {1988},
url = {https://doi.org/10.1016/0370-1573(88)90170-6}
}

@article{Baran05,
title = {Reaction dynamics and the nuclear symmetry energy},
author = {V. Baran and M. Colonna and V. Greco and M. Di Toro},
journal = {Phys. Rep.},
volume = {410},
pages = {335},
year = {2005},
url = {https://doi.org/10.1016/j.physrep.2004.12.004}
}

@article{Li08,
title = {Recent progress and new challenges in isospin physics with heavy-ion reactions},
author = {B. A. Li and L. W. Chen and C. M. Ko},
journal = {Phys. Rep.},
volume = {464},
pages = {113},
year = {2008},
url = {https://doi.org/10.1016/j.physrep.2008.04.005}
}

@article{Li02,
title = {Probing mechanical and chemical instabilities in neutron-rich matter},
author = {B. A. Li and Andrew T. Sustich and M. Tilley and B. Zhang},
journal = {Nucl. Phys. A},
volume = {699},
pages = {493},
year = {2002},
url = {https://doi.org/10.1016/S0375-9474(01)01291-X}
}

@article{Li95,
title = {Formation of superdense hadronic matter in high energy heavy-ion collisions},
author = {B. A. Li and C. M. Ko},
journal = {Phys. Rev. C},
volume = {52},
pages = {2037},
year = {1995},
url = {https://doi.org/10.1103/PhysRevC.52.2037}
}

@article{Lin05,
title = {Multiphase transport model for relativistic heavy ion collisions},
author = {Z. W. Lin and C. M. Ko and B. A. Li and B. Zhang and S. Pal},
journal = {Phys. Rev. C},
volume = {72},
pages = {064901},
year = {2005},
url = {https://doi.org/10.1103/PhysRevC.72.064901}
}

@book{PDG88,
title     = {Total Cross sections for Reactions of High Energy Particles},
editor    = {A. Baldini and V. Flaminio and W. G. Moorhead and D. R. O. Morrison},
publisher = {Springer-Verlag},
address   = {Berlin},
year      = {1988}
}

@article{Olive14,
title = {Review of Particle Physics},
author = {K. A. Olive and others},
collaboration = {Particle Data Group},
journal = {Chin. Phys. C},
volume = {38},
pages = {090001},
year = {2014},
url = {https://doi.org/10.1088/1674-1137/38/9/090001}
}

@article{Wei24,
title = {Kaon production in HADES experiment in Au + Au collisions at sNN = 2.4 GeV},
author = {G. F. Wei and Y. L. Zhao},
journal = {Phys. Rev. C},
volume = {110},
pages = {054615},
year = {2024},
url = {https://doi.org/10.1103/PhysRevC.110.054615}
}

@article{Long24,
titile = {Effects of incompressibility K0 in heavy-ion collisions at intermediate energies},
author = {X. X. Long and G. F. Wei},
journal = {Phys. Rev. C},
volume = {109},
pages = {054619},
year = {2024},
url = {https://doi.org/10.1103/PhysRevC.109.054619}
}

@article{FOPI12,
title = {Systematics of azimuthal asymmetries in heavy ion collisions in the 1A GeV regime},
author = {W. Reisdorf and others},
collaboration = {FOPI Collaboration},
journal = {Nucl. Phys. A},
volume = {876},
pages = {1-60},
year = {2012},
url = {https://doi.org/10.1016/j.nuclphysa.2011.12.006}
}

@article{Hillm20,
title = {First, second, third and fourth flow harmonics of deuterons and protons in Au +Au reactions at 1.23 AGeV},
author = {Paula Hillmann and Jan Steinheimer and Tom Reichert 
and Vincent Gaebel and Marcus Bleicher and Sukanya Sombun and Christoph Herold and Ayut Limphirat},
journal = {J. Phys. G: Nucl. Part. Phys.},
volume = {47},
pages = {055101},
year = {2020},
url = {https://doi.org/10.1088/1361-6471/ab6fcf}
}

@article{Nagle96,
title = {Coalescence of deuterons in relativistic heavy ion collisions},
author = {J. L. Nagle and B. S. Kumar and D. Kusnezov and H. Sorge and R. Mattiello},
journal = {Phys. Rev. C},
volume = {53},
pages = {367},
year = {1996},
url = {https://doi.org/10.1103/PhysRevC.53.367}
}

@article{Pan93,
title = {From sideward flow to nuclear compressibility},
author = {Q. B. Pan and P. Danielewicz},
journal = {Phys. Rev. Lett.},
volume = {70},
pages = {2062},
year = {1993},
url = {https://doi.org/10.1103/PhysRevLett.70.2062}
}

@article{Zhang94,
title = {Momentum-dependent nuclear mean fields and collective flow in heavy-ion collisions},
author = {J. M. Zhang and S. D. Gupta and C. Gale},
journal = {Phys. Rev. C},
volume = {50},
pages = {1617},
year = {1994},
url = {https://doi.org/10.1103/PhysRevC.50.1617}
}

@article{LiAng_2021,
title = {Constraints on the Maximum Mass of Neutron Stars with a Quark Core from GW170817 and NICER PSR J0030+0451 Data},
author = {Li, Ang and Miao, Zhi Qiang and Han, Sophia and Zhang, Bing},
journal = {Astrophys. J.},
volume = {913},
pages = {27},
year = {2021},
url = {https://doi.org/10.3847/1538-4357/abf355}
}

@article{Li_2026,
title = {Bayesian Inference of Hybrid Star Properties from Future High-precision Measurements of Their Radii},
author = {Li, Bao An and Grundler, Xavier and Xie, Wen Jie and Zhang, Nai Bo},
journal = {Astrophys. J.},
volume = {998},
pages = {262},
year = {2026},
url = {https://doi.org/10.3847/1538-4357/ae38c0}
}
\end{document}